%% file: main.tex
\documentclass[conference]{IEEEtran}
\usepackage{amsmath,amssymb,amsfonts}
\usepackage{algorithmic}
\usepackage{graphicx}
\usepackage{booktabs}
\usepackage{multirow}
\usepackage{amsmath}
\usepackage{textcomp}
\usepackage{xcolor}
\usepackage[sorting=none]{biblatex}
\begin{document}

\title{Estimation of Electronic Band Gap Energy From Material Properties Using Machine Learning}

\author{\IEEEauthorblockN{Sagar Prakash Barad\IEEEauthorrefmark{1},
Sajag Kumar\IEEEauthorrefmark{1} and Subhankar Mishra\IEEEauthorrefmark{2}}
\IEEEauthorblockA{\IEEEauthorrefmark{1,2}School of Physical Sciences, \IEEEauthorrefmark{2} School of Computer Sciences \\ 
National Institute of Science Education and Research, Bhubaneswar\\
An OCC of HBNI\\
Email: \{sagar.barad, sajag.kumar, smishra\}@niser.ac.in}}

\maketitle

\begin{abstract}

Machine learning techniques are utilized to estimate the electronic band gap energy and forecast the band gap category of materials based on experimentally quantifiable properties. The determination of band gap energy is critical for discerning various material properties, such as its metallic nature, and potential applications in electronic and optoelectronic devices. While numerical methods exist for computing band gap energy, they often entail high computational costs and have limitations in accuracy and scalability. A machine learning-driven model capable of swiftly predicting material band gap energy using easily obtainable experimental properties would offer a superior alternative to conventional density functional theory (DFT) methods. Our model does not require any preliminary DFT-based calculation or knowledge of the structure of the material. We present a scheme for improving the performance of simple regression and classification models by partitioning the dataset into multiple clusters. A new evaluation scheme for comparing the performance of ML-based models in material sciences involving both regression and classification tasks is introduced based on traditional evaluation metrics. It is shown that on this new evaluation metric, our method of clustering the dataset results in better performance.
\end{abstract}


\input{introduction}
\input{materials_methodology}
\input{results}
\input{conclusion}

\end{document}

%% file: introduction.tex
\section{Introduction}

The energy of electrons in materials can not be arbitrary. Electrons are allowed to have specific particular energy values. A band is a set of closely spaced energy values an electron can occupy [1]. The concept of the band gap refers to the energy level difference between two distinct bands within a material. Specifically, concerning band gap energy, it signifies the variance in energies between the valence and conduction bands. The conduction band encompasses allowable energy levels where electrons can conduct electricity within the material, while the valence band represents the permissible energies occupied by the outermost electrons of the material's atoms. Depending on their arrangement, the valence and conduction bands may demonstrate different characteristics: they may overlap (resulting in a zero band gap), indicating a metallic material, or they may have a slight separation (yielding a small band gap), suggestive of a semiconductor. Alternatively, they may be significantly apart (resulting in a wide band gap), indicating an insulator. The determination of band gap energy holds paramount importance for various purposes. For example, it illuminates the energy required to initiate conductivity in a semiconductor's electrons. Additionally, band gap energy plays a crucial role in assessing a material's potential for applications across electronic and optoelectronic devices [2]. Materials are divided into two broad categories on the basis of the kind of band gap, direct and indirect band gap materials. If the minima of the conduction band and the maxima of the valence band are on top of each other, the material is called direct band gap material. If they are separated, it is an indirect band gap material [1]. The directness of the band gap also plays an important role in determining the potential applications of materials in devices [3]. \par

Traditionally properties of materials such as the band gap energy are computationally calculated using density function theoretic methods [4]. Density functional theory (DFT) is used for computational studies of materials. But it is not able to predict the band gap energy of materials accurately [5]. However, the accuracy is good enough for most practical purposes. But DFT-based calculations for band gap energy are very time-consuming and require a lot of computational resources [6]. Improving DFT-based calculations for band gap estimation is still an active area of research, with many new techniques and approximations being explored to bring down the computational cost and increase the accuracy [7]. Band gap energy can also be determined experimentally, but even this is not easy if the band gap range is not known before. Usually, the experimental methods are tailor-made for materials with specific band gap ranges [8]. The method for analyzing data from such experiments is also quite complicated and not ideal for a first estimation [9]. \par

Recent advancements in machine learning for predicting band gaps focus on constructing models trained on a specific category of materials. Although these models excel at predicting band gaps for specialized materials, their effectiveness is somewhat limited as they don't need to learn from the ground up due to their specificity. They come pre-equipped with knowledge regarding the potential range of band gaps for the materials they encounter, thus predicting within this predetermined range. Ideally, it would be preferable to predict band gaps solely based on fundamental material properties. Additionally, certain models utilize features computed through density functional theory (DFT) [10]. These DFT-based preliminary calculations make using the model computationally expensive. Our objective is to develop a universal machine learning model capable of estimating the electronic band gap of materials solely relying on their fundamental properties. This model aims to bypass initial DFT-based computations and eliminate the necessity of prior knowledge regarding the material's structure. \par
Such a model is not expected to match the performance of composition and structure-based models, which are made for special classes of materials. But it can be used to predict the band gap energy and gap type of any material with relatively good accuracy without any preliminary DFT-based calculation or knowledge of its structure. Such a model will not only help in cheap band gap engineering but is also more interpretable than graph neural network-based models. The performance of the baseline model is improved upon by introducing a clustering step that partitions the dataset, and further models are trained on these individual clusters. The final output for a material belonging to a particular cluster is obtained from the models with the weights optimized on its corresponding cluster. This idea is discussed in detail in the following section.

%% file: materials_methodology.tex
\section{Materials and Methodology}
\begin{figure}
    \centering
    \begin{minipage}[b]{0.2\textwidth}
        \includegraphics[width=\textwidth]{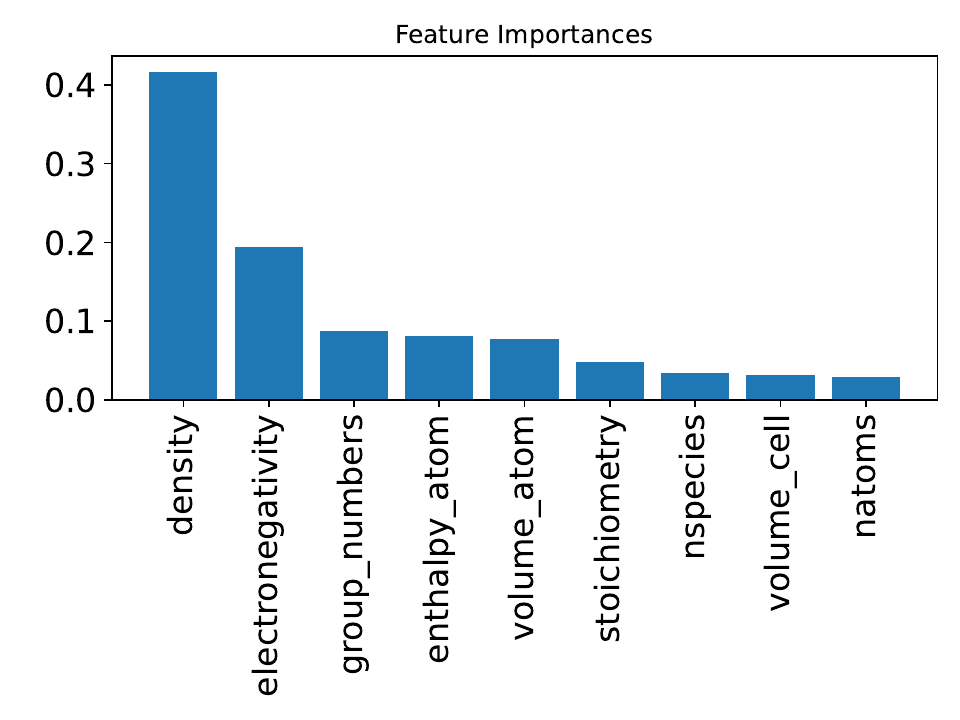}
        \caption{Feature importance plot for band gap regression}
        \label{fig:bandgap_feature_importance}
    \end{minipage}
    \hfill
    \begin{minipage}[b]{0.2\textwidth}
        \includegraphics[width=\textwidth]{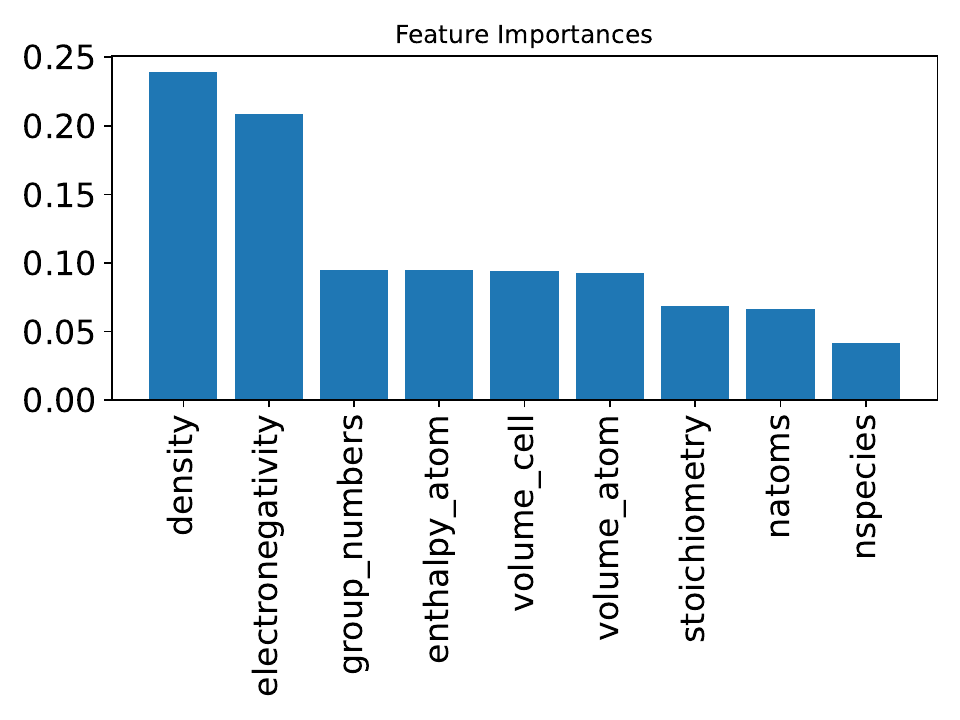}
        \caption{Feature importance plot for gap type classification}
        \label{fig:gaptype_feature_importance}
    \end{minipage}
\end{figure}

\begin{figure}
    \centering
    \includegraphics[width = 0.35\textwidth]{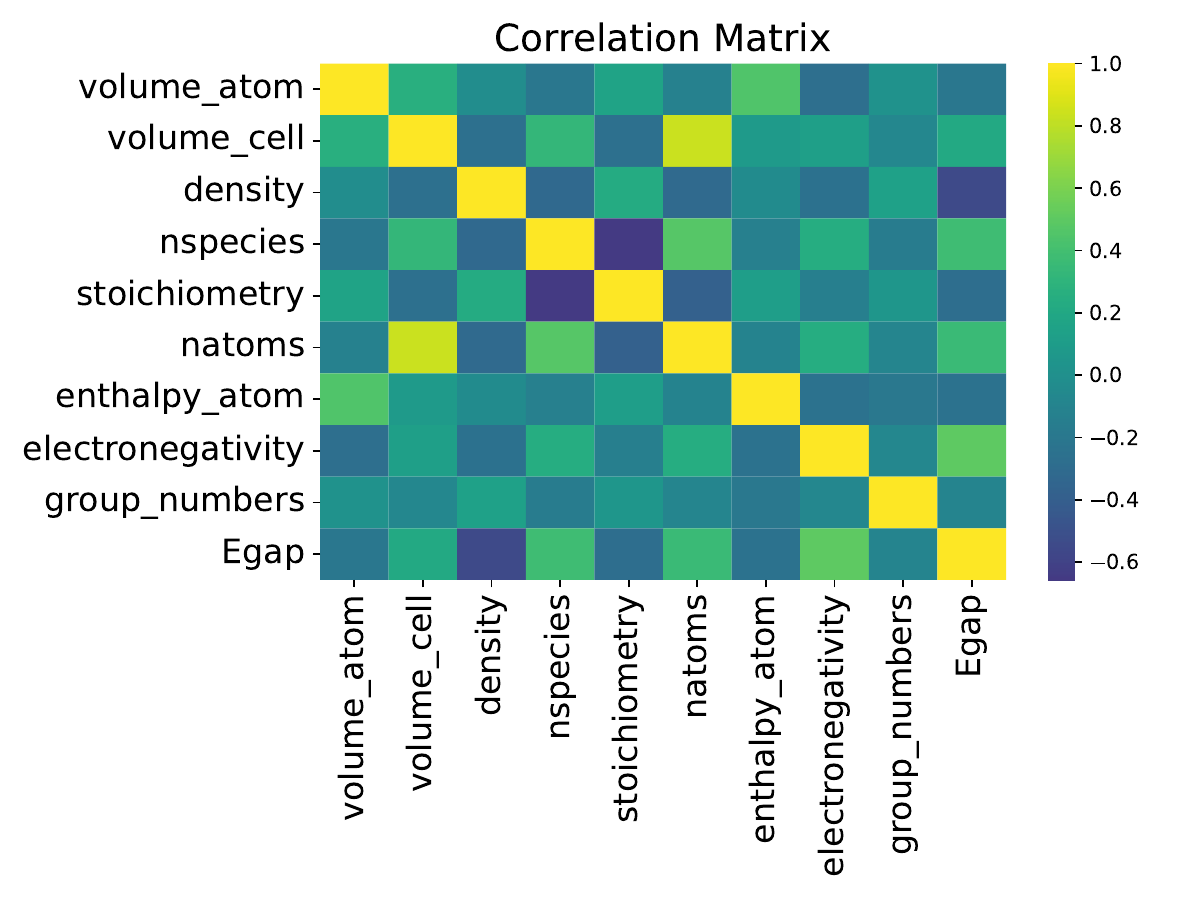}
    \caption{Correlation matrix for the dataset.}
    \label{fig:corr}
\end{figure}

We selected a benchmark dataset from Benchmark AFLOW Data Sets for Machine Learning [11] and chose specific features to train our machine learning models to predict bandgap and gap type. The selected features are shown in \ref{tab:features}. These features are chosen because they are easily determined from elementary experiments for a new material. They do not require any DFT calculations or knowledge of three dimensional structure of the material. The features \emph{electronegativity} and \emph{group\_numbers} are engineered by us, for \emph{group\_numbers} the elements in species are replaced with their corresponding group numbers, and for \emph{electronegativity} the average of the electronegativity values of the constituent atoms referenced using the Pauling scale is taken. The selected features capture important information about the composition of the material and its electronic structure, which are known to have a significant impact on the bandgap. For example, smaller values of \emph{volume\_cell} and \emph{volume\_atom} tend to have higher band gap due to increased electron confinement, while larger values tend to have lower band gap due to reduced electron confinement, \emph{Natoms} can affect the band gap energy via the density of states, and materials with high \emph{electronegativity} differences between constituent atoms tend to have larger band gap due to increased ionic bonding. By selecting these specific features, the aim is to capture the most critical information about the material's composition and electronic properties that are known to influence the band gap and are easily determinable. The feature importance of these features for band gap regression and gap type classification is shown in figure \ref{fig:bandgap_feature_importance} and figure \ref{fig:gaptype_feature_importance} respectively. The correlation matrix for the dataset is shown in figure \ref{fig:corr}.

\begin{table}[htp]
\centering
\caption{The selected features and their definition.}
\begin{tabular}{@{}ll@{}}
\toprule
\textbf{Feature Name} & \textbf{Definition}                                 \\ \midrule
Egap                  & The electronic band gap energy.                     \\
group\_numbers        & Group numbers of the atoms present in the material. \\
electronegativity     & Engineered using the Pauling scale.                 \\
enthalpy\_atom        & Enthalpy of atoms present in the material.          \\
natoms                & Number of atoms in one molecule of the material.    \\
stoichiometry         & Stoichiometry of the material.                      \\
nspecies              & Number of species of atoms present in the material. \\
density               & Density of the material.                            \\
volume\_cell          & Volume of the unit cell.                            \\
volume\_atom          & Volume of the atoms present in the unit cell.       \\ \bottomrule
\end{tabular}

\label{tab:features}
\end{table}

We used ordinal encoding to represent the data in the stoichiometry column and group numbers as numerical arrays while preserving their order. After encoding, the dataset was standardized and normalized to ensure that all features have a similar range and distribution, which improves the efficiency and performance of the machine learning models. These preprocessing steps helped us to incorporate categorical data into our models and ensure that they were well-conditioned and ready for training.
The compiled dataset with 55,298 samples and 9 features is split into two parts, a training set with 52,534 samples and a test set with 2,765 samples. All machine learning models are trained on the training set. The test set is kept separate until the final evaluation. By doing so, it is ensured that our models were not exposed to the test set during training and were evaluated on previously unseen data.\par

The training set has 27396 metals and 25138 non-metals. Among the non-metals 15838 are indirect band gap type while 9300 are direct band gap type. The maximum value of the band gap is 9.0662 eV. The average band gap is 1.3176 eV and the standard deviation is 1.8079 eV.\par

The test set has 1446 metals and 1319 non-metals. Among the non-metals 843 are indirect band gap type while 476 are direct band gap type. The maximum value of the band gap is 9.084 eV. The average band gap is 1.3154 eV and the standard deviation is 1.8219 eV.

\subsection{Machine Learning Algorithms}

Random forest and gradient boosted trees have been used for most of the classification and regression tasks. The k-means algorithm is used for clustering. Apart from these XGBoost has also been used for some of the regression tasks. These are very common machine learning algorithms that are widely used. 

\subsection{Architectures}

\begin{figure}[htp]
    \centering
    \includegraphics[width = 0.35\textwidth]{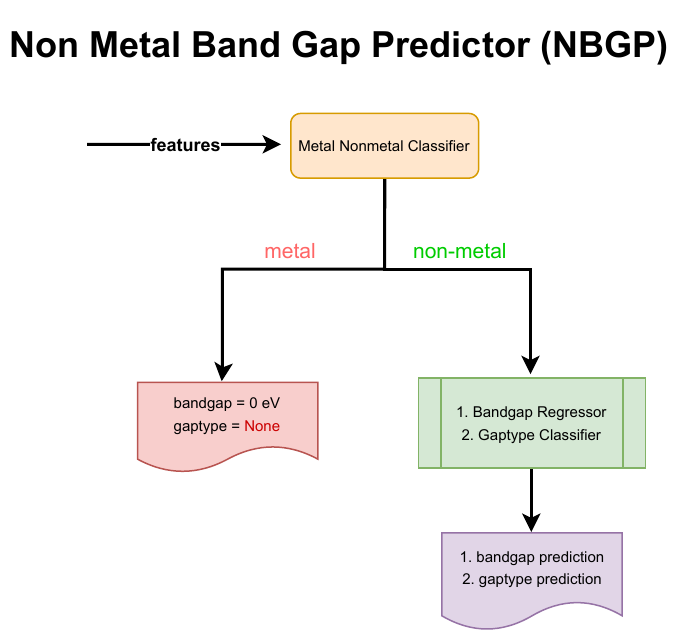}
    \caption{The first architecture without the utilisation of clustering on non-metals.}
    \label{fig:arch1}
\end{figure}

\begin{figure}[htp]
    \centering
    \includegraphics[width = 0.35\textwidth]{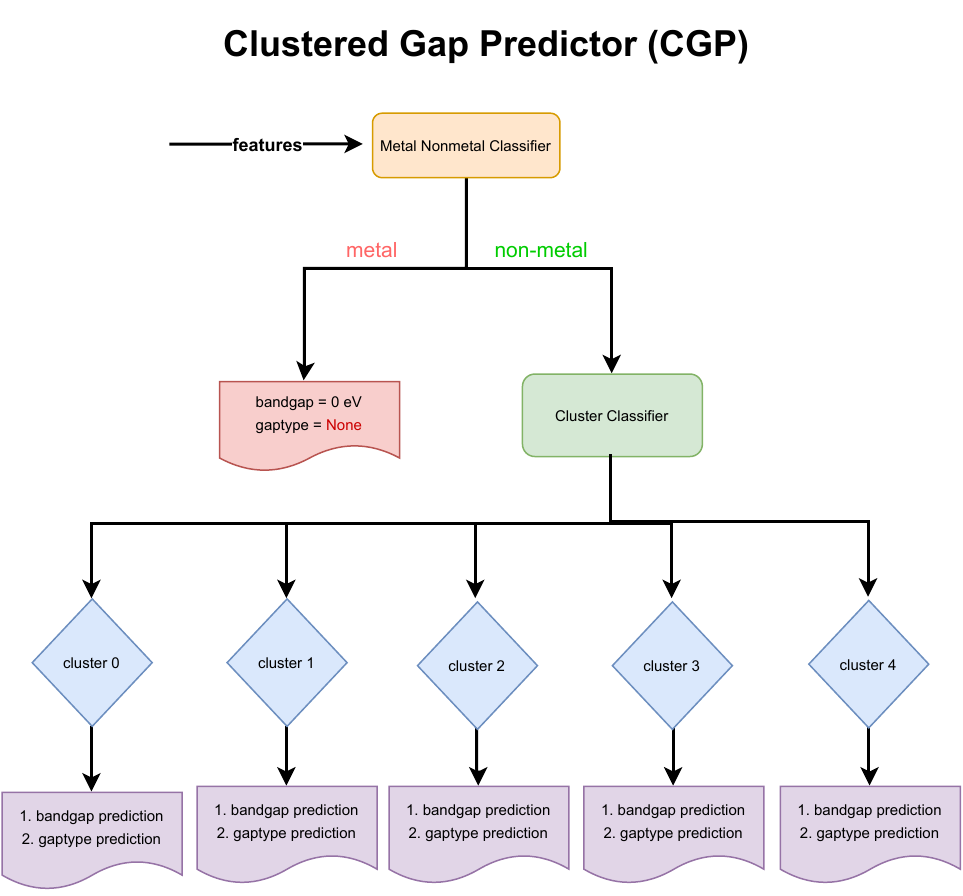}
    \caption{The second architecture with clustering on non-metals. We call this the clustered gap predictor (CGP).}
    \label{fig:arch2}
\end{figure}

Two architectures are built for estimating the band gap and its type. The first architecture is shown in figure \ref{fig:arch1} and the second architecture is shown in figure \ref{fig:arch2}. The metal - non-metal classifier is same for both the models. The only difference is the utilisation of clustering on non-metals in the second architecture. While in the first architecture algorithms for band gap and gap type prediction are trained on all of the non-metals in the second architecture, firstly the non-metals are clustered in five clusters and models are trained for band gap and gap type prediction on each of the clusters separately.\par
It is expected that upon clustering the performance of the model will improve. Because similar materials would be clustered together resulting in better performance of the algorithms trained on the clusters and accordingly better net performance. Stacking models together in any of the architecture has been avoided because it was observed that models with good performance when stacked together performed worse than the worst performing model. 

\subsection{Evaluation Metrics}

Our model predicts three things. It predicts whether a new material is a metal; if it is not, then it estimates its band gap and predicts the band gap type. These predictions are outputs of two different types of machine learning algorithms. The band gap prediction is the output of a regression algorithm; it takes real values, while the whether the material is metal and band gap type are the outputs of classification algorithms, which take binary values. We define the following score to evaluate the performance of different architectures for this task.

\begin{equation}
    \begin{split}\label{eq:eval}
    \text{Score} =&(0.3)(F1_{MNM})(1-P_{MNM}) + \\ &(0.4)\left(1- \frac{MAE_{Egap}}{Egap_{max}} \right) + \\ &(0.3)(F1_{GT})
\end{split}
\end{equation}

Where $F1_{MNM}$ is the F1 score of the metal - non-metal classifier, $P_{MNM}$ is the precision of the metal - non-metal classifier, $MAE_{Egap}$ is the mean absolute error in band gap estimation, $Egap_{max}$ is the maximum value of band gap in the dataset and $F1_{GT}$ is the F1 score of the gap type classifier.\par

The score ranges from 0 to 1, where a score of 1 indicates perfect performance and a score of 0 indicates the worst possible performance. This evaluation metric assesses the performance of predicting metal vs. non-metal, as well as the regression of band gap values and the classification of band gap types. In addition, it penalizes the misclassification of samples as metals and evaluates the performance of the gap type classifier and band gap regressor for such samples. Penalizing the misclassification of non-metals as metals heavily is important. If a non-metal is misclassified as metal, all three predictions are definitely wrong. But in case metal is misclassified as non-metal, it still has some hope of being assigned a very small band gap.

%% file: results.tex
\section{Results}

\begin{figure}[htp]
    \centering
    \includegraphics[width=0.35\textwidth]{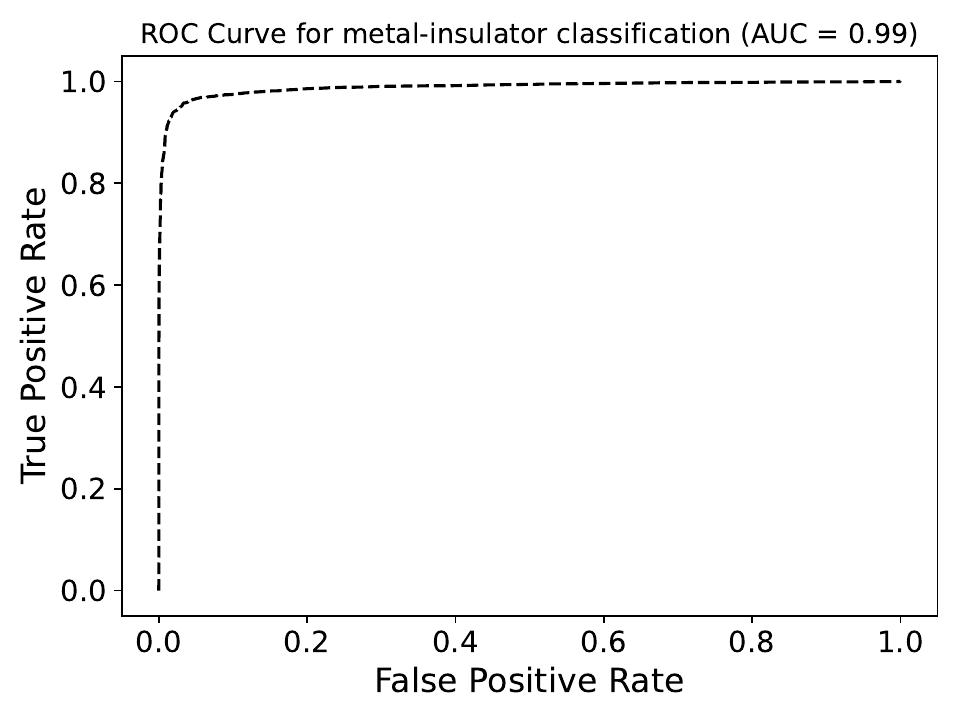}
    \caption{AUC-ROC curve for metal - non-metal classification.}
    \label{fig:mnm}
\end{figure}

The first step in our architecture is to classify materials as metals or non-metals. It is a critical step for incorrect labeling of materials will lead to absolutely incorrect predictions. For example, if a non-metal is misclassified as a metal, it will be assigned a band gap of zero and gap type of \emph{None}. A good metal classifier is thus necessary for the good performance of a model. In our case, we found that a gradient-boosting classifier (GBC) worked the best for this task. Our classifier has an AUC score of 0.99, an AUC-ROC curve shown in figure \ref{fig:mnm}, which ensures minimal misclassifications. This step is the same for the two architectures that have been built.\par

\begin{table}[htp]
\centering
\caption{Summary of results of various algorithms for band gap estimation.}
\begin{tabular}{@{}ccc@{}}
\toprule
\textbf{Model} & \textbf{MAE} & \textbf{$R^2$} \\ \midrule
RFR            & 0.373        & 0.833          \\
GBR            & 0.396        & 0.839          \\
XGB            & 0.416        & 0.834          \\ \bottomrule
\end{tabular}
\bigskip

\label{tab:regression}
\end{table}

\begin{figure}[htp]
    \centering
    \includegraphics[width=0.35\textwidth]{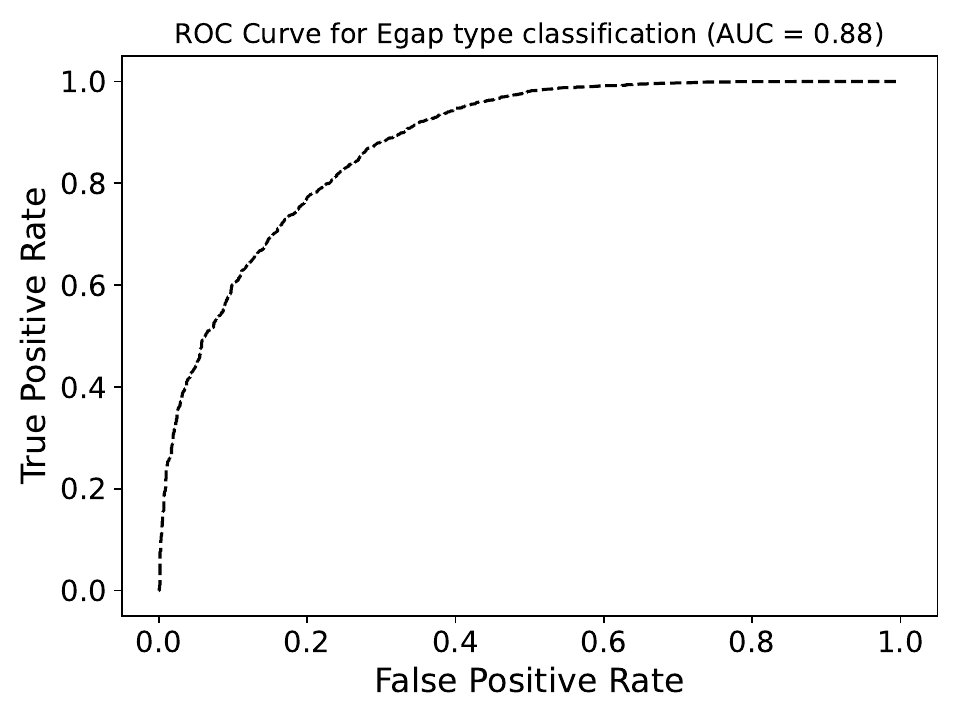}
    \caption{AUC-ROC curve for band gap type classification.}
    \label{fig:type}
\end{figure}

If a material is classified as a non-metal, the next step is to estimate the band gap and predict its type (direct or indirect). For the estimation of the band gap, three algorithms, random forest regressor (RFR), gradient boosting regressor (GBR), and XGBoost (XGB), gave similar results. With GBR performing the best in terms of $R^2$ score and RFR performing the best in terms of the MAE. The ensemble of these models gave worse results. The results of these algorithms are summarised in table \ref{tab:regression}. These algorithms are only trained on the non-metals in the dataset. On the non-metals in the dataset, a band gap type classifier is also trained. A GBC worked the best for this task as well. The AUC-ROC curve for the classifier is shown in figure \ref{fig:type}. An AUC score of 0.88 was achieved for this classification task.\par  

\begin{table*}[htp]
\centering
\caption{Statistics of the obtained clusters. Here, $\mu$ is the average band gap, and $\sigma$ is the standard deviation of the band gap of the materials in the corresponding clusters in eV.}
\begin{tabular}{@{}cccccc@{}}
\toprule
\multirow{2}{*}{\textbf{Cluster Number}} & \multirow{2}{*}{\textbf{Number of Samples}} & \multicolumn{2}{c}{\textbf{gap\_type Frequencies}} & \multicolumn{2}{c}{\textbf{E\_gap}} \\ \cmidrule(l){3-6} 
                                         &                                             & \textbf{Direct}         & \textbf{Indirect}        & \textbf{$\mu$}  & \textbf{$\sigma$} \\ \cmidrule(r){1-2}
0                                        & 8546                                        & 0.368125                & 0.631875                 & 2.63            & 1.81              \\
1                                        & 4764                                        & 0.355374                & 0.644626                 & 2.75            & 1.56              \\
2                                        & 5048                                        & 0.414025                & 0.585975                 & 2.34            & 1.61              \\
3                                        & 3483                                        & 0.383003                & 0.616997                 & 3.25            & 1.61              \\
4                                        & 3297                                        & 0.314528                & 0.685472                 & 3.19            & 1.57              \\ \bottomrule
\end{tabular}
\bigskip

\label{tab:clusters}
\end{table*}

The above algorithms for band gap estimation and gap type classification were trained on all the non-metals in the dataset. Now we cluster the non-metals into five different clusters and train models for band gap estimation and gap type classification on each cluster separately. The k-Means clustering algorithm is used for clustering the non-metals. The statistics of the five clusters obtained are shown in table \ref{tab:clusters}. Cluster 0 has the most number of samples, whereas cluster 4 has the least number of samples. The relative frequencies of direct and indirect band gap materials in each cluster are representative of their relative frequency in the whole dataset. The average value of the band gap is lowest in cluster 2 and highest in cluster 3. The spread of the band gap quantified through its standard deviation is lowest in cluster 1 and highest in cluster 0. However, the standard deviation is quite high for each of the clusters. It would have been ideal if materials with similar band gaps were clustered together, resulting in low standard deviation, but as the standard deviation of the dataset is itself quite high, this is the best result that one may obtain. \par

\begin{figure}[htp]
    \centering
    \includegraphics[width=0.35\textwidth]{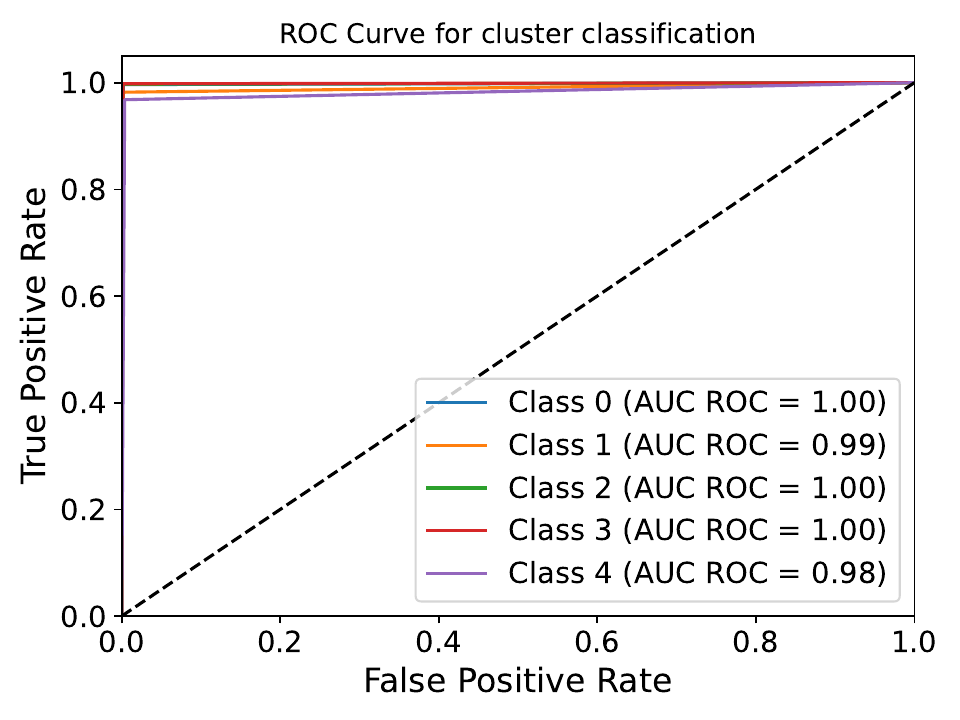}
    \caption{AUC-ROC curve for assigning new materials to their corresponding clusters.}
    \label{fig:clusters}
\end{figure}

Once the clusters are obtained, before training models on each one of them for band gap and gap type prediction, we need a method to assign new materials to the cluster that has similar materials. To do this multi-class classifier is trained. The AUC-ROC curve for each cluster is shown in figure \ref{fig:clusters}. A perfect AUC score is achieved for clusters 0, 2 and 3, an AUC score of 0.99 and 0.98 is achieved for clusters 3 and 4. This very good performance of the cluster classifier is extremely important for the clustered gap predictor (CGP). The key idea of training different models on different clusters and using these models trained on different clusters to make predictions works only if the mechanism of assigning new materials to these clusters works well. Just like the metal - non-metal classifier, the cluster classifier is also very important for an architecture with a good score. A random forest classifier (RFC) was found to perform the best for this task of cluster assignment.\par

\begin{figure}[htp]
    \centering
    \includegraphics[width=0.35\textwidth]{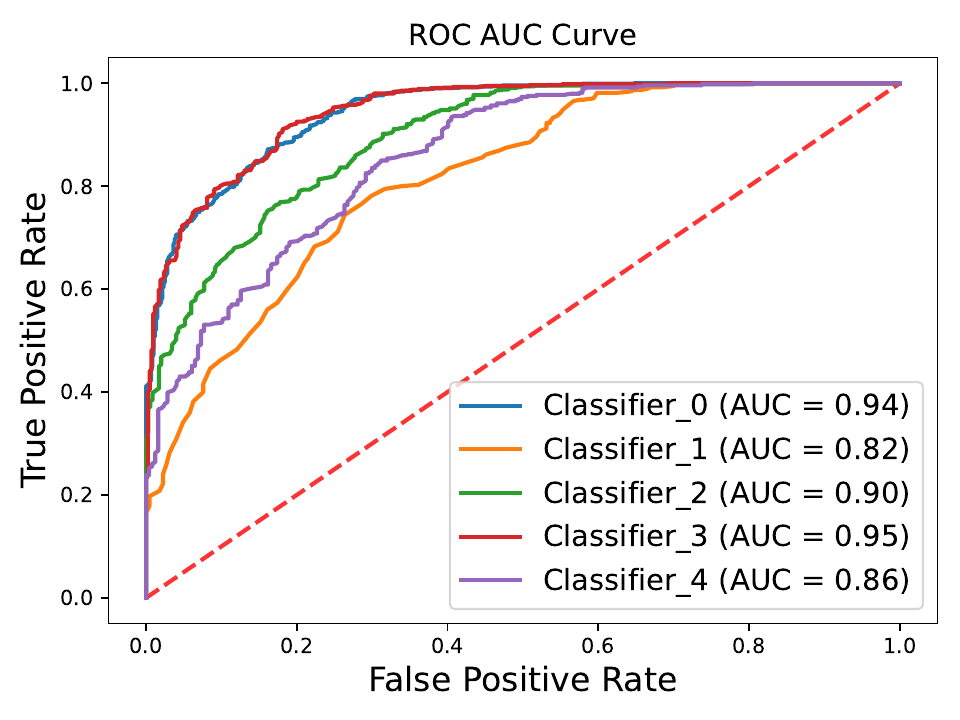}
    \caption{AUC-ROC curve for the band gap type classifiers on each cluster.}
    \label{fig:type_clusters}
\end{figure}

\begin{table}[htp]
\centering
\caption{Summary of results for band gap estimation on each cluster.}
\begin{tabular}{@{}cccc@{}}
\toprule
\textbf{Cluster}   & \textbf{Model} & \textbf{MAE} & \textbf{$R^2$} \\ \midrule
\multirow{2}{*}{0} & RFR            & 0.3          & 0.909          \\
                   & XGB           & 0.265        & 0.917          \\
\multirow{2}{*}{1} & RFR            & 0.483        & 0.706          \\
                   & XGB           & 0.448        & 0.722          \\
\multirow{2}{*}{2} & RFR            & 0.294        & 0.899          \\
                   & XGB           & 0.274        & 0.902          \\
\multirow{2}{*}{3} & RFR            & 0.538        & 0.782          \\
                   & XGB           & 0.51         & 0.785          \\
\multirow{2}{*}{4} & RFR            & 0.536        & 0.709          \\
                   & XGB           & 0.508        & 0.719          \\ \bottomrule
\end{tabular}
\bigskip

\label{tab:reg_clusters}
\end{table}

Once this a mechanism for assigning new materials to their corresponding clusters is in place, models for band gap and band gap tyep classification are trained on individual clusters. Random forest regressor (RFR) and XGBoost (XGB) performed the best for the band gap estimation. The best $R^2$ score is obtained for cluster 0 with XGB, while the worst $R^2$ score is obtained for cluster 1 with RFR. The least MAE is obtained for cluster 0 with XGB, while the largest MAE is obtained for cluster 3 with RFR. In general, for all the clusters, XGB performed slightly better than RFR in terms of both MAE and $R^2$. The good performance of cluster 0 is due to its large size. While the bad performance of cluster 4 is due to its small size. The performance of cluster 1 is slightly puzzling. It is larger than clusters 3 and 4 and has the least value of standard deviation among all clusters, so it is expected to have better performance than at least clusters 3 and 4, but it performs worse than cluster 3 for both the algorithms and worse than cluster 4 for RFR. It may seem here that only two clusters have shown better performance than the regression model on the non-metals, but these two clusters contain more materials than the other three combined, so the net result is indeed better. The results of the regression models on the clusters is summarised in table \ref{tab:reg_clusters}. Gap type classifiers are also trained on each of the clusters. Random forest classifiers performed the best for this task on each of the clusters. The AUC-ROC curves for each cluster are shown in figure \ref{fig:type_clusters}. Once again, the best performance is obtained on cluster 0, but the worst is obtained on cluster 1. The gap type classifier on non-metals had an AUC score of 0.88; after clustering, three of the five clusters have an AUC score greater than 0.88, resulting in a net increase. \par

\begin{figure}[htp]
    \centering
    \begin{minipage}[b]{0.23\textwidth}
        \includegraphics[width=\textwidth]{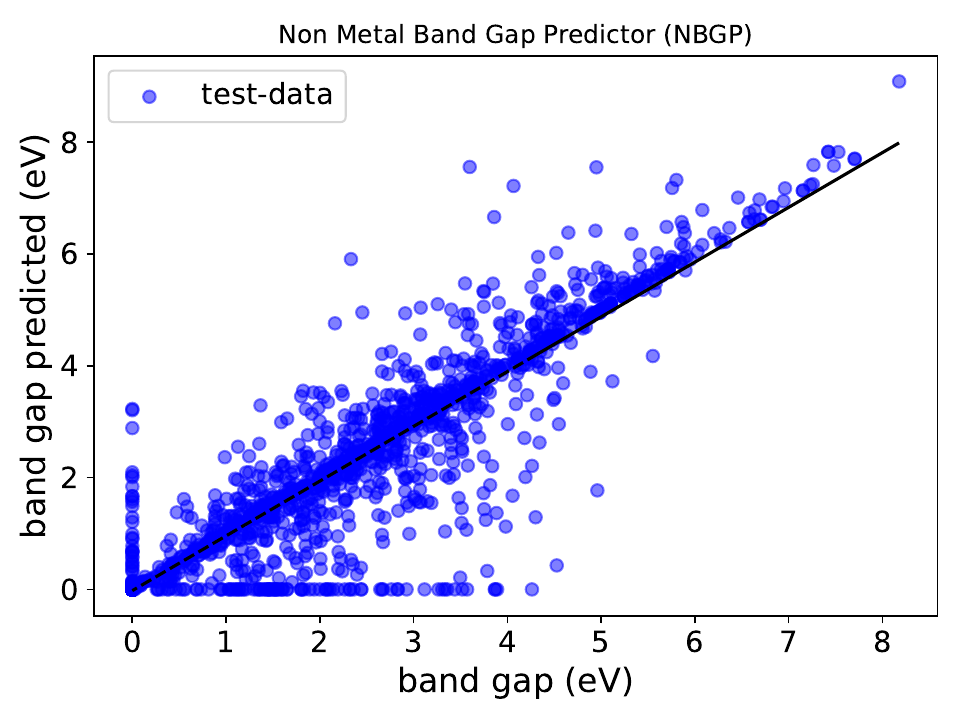}
        \caption{Prediction performance of the first architecture.}
        \label{fig:nmgp}
    \end{minipage}
    \hfill
    \begin{minipage}[b]{0.23\textwidth}
        \includegraphics[width=\textwidth]{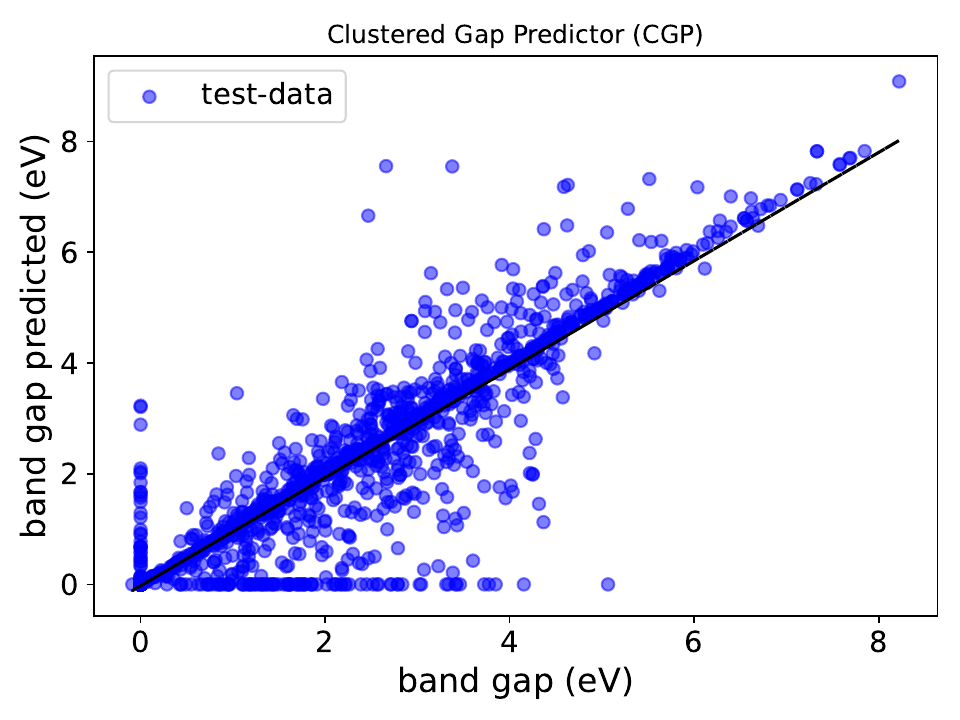}
        \caption{Prediction performance of the clustered gap predictor.}
        \label{fig:cgp}
    \end{minipage}
\end{figure}

\begin{table*}[htp]
\centering
\caption{Summary of results for architectures with and without clustering.}
\begin{tabular}{@{}clcccccccl@{}}
\toprule
\multicolumn{2}{c}{\multirow{2}{*}{\textbf{Model}}}                                         & \multicolumn{2}{c}{\textbf{\begin{tabular}[c]{@{}c@{}}Metal\\ Classification\end{tabular}}} & \multicolumn{2}{c}{\textbf{\begin{tabular}[c]{@{}c@{}}Band Gap\\ Prediction\end{tabular}}} & \multicolumn{2}{c}{\textbf{\begin{tabular}[c]{@{}c@{}}Gap Type\\ Classification\end{tabular}}} & \multicolumn{2}{c}{\multirow{2}{*}{\textbf{Score}}} \\ \cmidrule(lr){3-8}
\multicolumn{2}{c}{}                                                                        & \textbf{Accuracy}                            & \textbf{F1 Score}                            & \textbf{$R^2$}                                & \textbf{MAE}                               & \textbf{Accuracy}                              & \textbf{F1 Score}                             & \multicolumn{2}{c}{}                                \\ \cmidrule(r){1-2} \cmidrule(l){9-10} 
\multicolumn{2}{c}{\begin{tabular}[c]{@{}c@{}}Without\\ Clustering\end{tabular}}            & 0.9508                                       & 0.9521                                       & 0.8906                                        & 0.2496                                     & 0.8665                                         & 0.8659                                        & \multicolumn{2}{c}{0.9299}                          \\
\multicolumn{2}{c}{\begin{tabular}[c]{@{}c@{}}Clustered Gap\\ Predictor (CGP)\end{tabular}} & 0.9508                                       & 0.9521                                       & 0.8930                                         & 0.2321                                     & 0.8770                                          & 0.8769                                        & \multicolumn{2}{c}{0.9336}                          \\ \bottomrule
\end{tabular}
\bigskip

\label{tab:final}
\end{table*}

The final results for the two architectures are summarised in table \ref{tab:final}. The score for the architectures is computed using equation \ref{eq:eval}. It can be seen that the clustered gap predictor has a better overall score, a better $R^2$ score and MAE for band gap estimation, and a better accuracy and F1 score for gap type prediction. The performance of both the architectures is visualised in figure \ref{fig:nmgp} and \ref{fig:cgp} .

%% file: conclusion.tex
\section{Conclusion}

In conclusion, we have shown that it is possible to build machine learning models based on easily determinable material properties for band gap estimation. Models that use features involving preliminary DFT-based calculations or encode the entire three-dimensional structure of the material perform better. But these models also need more resources and are less interpretable. It is also demonstrated that training different models on different clusters results in better performance. \par 
The clustered gap predictor (CGP) is trained on a big dataset with 52534 samples. It classifies materials into metals and non-metals and predicts the band gap and gap type of non-metals. Gradient boosted classifier was used to train the metal-non-metal classifier. It has an AUC score of 0.99. Random forest regressor and XGBoost were trained for band gap estimation. XGB performed better. The weighted average of the MAE on each cluster for band gap estimation is 0.2321 eV. Random forest classifiers were trained on each cluster for gap-type prediction. The AUC score on each cluster is greater than 0.97. CGP obtained a score of 0.9336 using the evaluation metric defined in equation \ref{eq:eval}.\par

Ensemble learning is not used because the ensembled models performed worse than the worst-performing model in regression tasks. A possible reason for this is high correlation among the trained models. Different models gave similar predictions for the same materials. In these situations, ensemble methods tend not to do well. An anomaly is the bad performance of regression algorithms on cluster 1, which has a considerable amount of samples (4764) and the least spread (quantified by the standard deviation) in band gap values throughout the samples in the cluster.\par

Future works in this direction should focus on developing a better clustering algorithm. At present, the number of clusters is a parameter entered manually. Ideally similar clusters should emerge without specifying the number of clusters beforehand. A challenge that needs to be overcome in this direction is ensuring that the clusters that appear have sufficient samples for training machine learning models. A larger dataset for training is also desirable. It has been observed that clusters with large number of samples performed much better than clusters with smaller number of samples. If each cluster could be assigned a large number of examples, the performance would significantly increase. With larger datasets, one can also explore using neural networks for regression and classification tasks. We have only focused on the band gap of materials in this work. Similar ideas may be employed for predicting other material properties, such as their conductivity, mobility, and formation energy, which are essential for device applications,.

%% file: main.bbl
\begin{thebibliography}{00}
\bibitem{b1} Ashcroft and Mermin. ``Solid State Physics''. Harcourt
College Publishers.
\bibitem{b2} Amit Sitt, Ido hadar, and Uri Banin. ``Band-gap engineering, optoelectronic properties and applications of
colloidal heterostructured semiconductor nanorods''. Nanotoday (2013).
\bibitem{b3} Jialin Yang et al. ``Recent advances in optoelectronic and microelectronic devices based on ultrawidebandgap semiconductors''. Progress in Quantum Electronics (2022).
\bibitem{b4} David S. Sholl and Janice A. Steckel. ``Density Functional Theory: A Practical Introduction''. Wiley, 2011.
\bibitem{b5} John P. Perdew and Mel Levy. ``Physical Content of
the Exact Kohn-Sham Orbital Energies: Band Gaps and
Derivative Discontinuities''. Physical Review Letters
(1983).
\bibitem{b6} John P. Perdew. ``Density functional theory and the band
gap problem''. Quantum Chemistry (1985)
\bibitem{b7} Pedro Borlido et al. ``Exchange-correlation functionals
for band gaps of solids: benchmark, reparametrization
and machine learning''. npj Computational Materials
(2020).
\bibitem{b8} Patrycja Makula, Michal Pacia, and Wojciech Macyk.
``How To Correctly Determine the Band Gap Energy of
Modified Semiconductor Photocatalysts Based on UV–
Vis Spectra''. The Journal of Physical Chemistry
Letters (2018).
\bibitem{b9} Brian D. Viezbicke et al. ``Evaluation of the Tauc
method for optical absorption edge determination: ZnO
thin films as a model system''. Basic Solid State
Physics (2015).
\bibitem{b10} Yu Zhang et al. ``Bandgap prediction of twodimensional materials using machine learning''. PLoS ONE (2019).
\bibitem{b11} Conrad L. Clement, Steven K. Kauwe, and Taylor D.
Sparks. ``Benchmark AFLOW Data Sets for Machine
Learning''. Integrating Materials and Manufacturing
Innovation (2020).
\end{thebibliography}
